# Collective plasmonic modes of metal nano-particles in two-dimensional periodic regular arrays

Yu-Rong Zhen , Kin Hung Fung, C. T. CHAN
*Physics Department, Hong Kong University of Science and Technology,*
*Clear Water Bay, Kowloon, Hong Kong*

We investigate the collective plasmonic modes of metal nano-particles in periodic two-dimensional (2D) arrays within a point-dipole description. As an open system, the full-dynamic dispersion relations of the 2D arrays are obtained through an efficient method which gives an effective polarizability describing the collective response of a system. Both the dispersion relations and mode qualities are simultaneously related to the imaginary part of the effective polarizability, which has contributions from the single-particle response as well as the inter-particle coupling. The transversal long-range dipolar interaction is dominated by a wave term together with a purely geometrical constant representing the static geometrical contribution to resonant frequencies. As concrete examples, we considered small Ag spheres arranged in a square lattice. We find that inside the light-cone, the transverse quasi-mode has a reasonably high mode quality while the two in-plane modes show significant radiation damping. Near the light-line, we observe strong coupling with free photons for the bands of the transverse mode and the transversal in-plane mode, and the longitudinal in-plane mode exhibits a negative group-velocity inside the light-cone. Vanishing group velocities in the light-cone for all the quasi-modes are found to be intrinsic properties of the 2D metal nano-sphere dense arrays.

**I. INTRODUCTION**

Recently the optical properties of clusters of metal nano-particles have attracted many interests because of their plausible nano-photonic applications such as the optical waveguides[1,2,3,4,5], biosensors[6,7], and sub-wavelength imaging[8,9]. The plasmonic resonance in noble metal particles, particularly nano-spheres, is often modeled as dipolar resonance as long as the interparticle spacing is large enough so that the high order resonances are no important. In those systems, the collective behavior of a cluster of particles can be directly modeled with coupled-dipole equations. Based on the coupled-dipole model, the plasmonic modes of metal nano-particles in one-dimensional periodic systems, such as single chain[2, 8,10,11,12,13,14,15] and double chains[15], have been extensively investigated. At the same time, there are also some interesting experimental[16,17] and theoretical[9,18,19,20,21] results on the two-dimensional arrays of nanoparticles or dipolar scatters.

A popular approach to treat those systems is to obtain the eigenmodes, i.e. the dispersion ($\omega$-$k$) relations, and for a true eigenmode description, one assumes that the system under investigation is a closed system (no coupling to the free photons and dissipative environment). Once conventional time-dependence $e^{-i\omega t}$ is adopted, in the absence of driving field, the oscillating frequencies $\omega$ of an actual system is a complex quantity with a imaginary part satisfying $\text{Im}(\omega) \leq 0$, because as an actual system it may be openly coupled to either dissipative environment or free photons, leading to decaying oscillation in time. As a result of the negative imaginary part of $\omega$, there is a divergence in the lattice sum of dynamic dipolar Green's function[11]. Usually there are three methods to avoid such difficulty. The first method is to model the system from "open" to "closed", in the quasi-static approximation[2,11] where $\text{Im}(\omega) = 0$. Thus the dispersion relations can be obtained in closed-form in 1D. The second approach is to consider a finite number of particles to emulate an infinite periodic structure[11], and dispersion relations are obtained by searching the loci of global minima of the determinant of a coupling matrix. The third approach is to evaluate the sums in the $\text{Im}(\omega) \geq 0$ half-plane and then carry out analytic continuation to the $\text{Im}(\omega) \leq 0$ half-plane[14,22]. The above approaches are quite successful in 1D system but they have their own limitations. The quasi-static approximation cannot account for the coupling between the free-propagating wave in the ambient medium and the eigenmodes of the system. The finite-size treatment in one-dimensional chain[11] requires the evaluation of the determinant of the coupling matrix, whose dimension N is proportional to the number of particles, as a function of the complex frequency at a given spatial wave vector. Further the pole-searching is not direct but should be accompanied with a tracing from the quasistatic limit because the number of poles is larger than the number of dipoles[14]. The evaluation of determinant typically requires an $O(N^3)$ algorithm and $O(N)$ times of evaluation are needed to find out the N poles of the complex function. The finite-size treatment has a complexity of $O(N^4)$ and becomes computationally demanding in two-dimensional arrays because of increased dimension of the coupling matrix. Lastly, while the analytic continuation method is realizable in infinite 1D chain, it becomes complicated in infinite 2D arrays. The above difficulties may explain the lack of full study of dispersion relations for the collective modes of periodic dipoles in infinite 2D arrays, while there are many such studies for 1D chains.

More efficient approaches are in fact available to obtain the dispersion relations and quality-factors simultaneously[15], and have been applied to 1D systems. For an open system, it is more appropriate (both from a computational point of view and from the physics point of view) to consider the system as a driven system instead of a system without external driving. In the presence of incident wave, an effective polarizability can be introduced within the framework of spectral decomposition[23,24] to describe the response from not only material properties but also collective coupling in a specific geometrical structure. We will apply such an approach to two-dimensional dipole arrays.

In this paper, we will focus on plasmonic dispersions of nano-spheres in infinite 2D periodic arrays. In Sec. II we retain in the point-dipole model which has been shown to give a very good description of small metal particles[25] as long as $a \leq d/3$, where $a$ is the sphere's radius and $d$ the lattice constant. We first discuss the dispersion relations within the quasi-static dipolar approximation in Sec. III. In Sec IV, we show the full-dynamic dispersion relations and quality factors simultaneously, in which the retardation effect is also included. Numerical results are given for silver nano-spheres arranged in infinite square lattice, followed by a discussion and a brief conclusion.

**II. COUPLED-DIPOLE EQUATIONS**

The local electromagnetic field at the position of a dipole consists of the incident field and the field radiated by all the other dipoles. On the local dipole the local field $\mathbf{E}_{local}$ induces a dipole moment $\mathbf{p}_i$, which is itself a radiation source to other dipoles. If the time-dependence is assumed to be $e^{-i\omega t}$ and the polarizability $\alpha$ is assumed to be isotropic and identical for every dipole, then for $i$ th dipole, we have the following coupled dipole equation,

$$\mathbf{p}_i = \alpha \left[ \mathbf{E}_{inc,i} + \sum_{j \neq i} \mathbf{G}(\mathbf{R}_i, \mathbf{R}_j) \mathbf{p}_j \right], \quad (2.1)$$

where

$$\mathbf{G}(\mathbf{R}_i, \mathbf{R}_j) \mathbf{p}_j = \left[ \left(1 - \frac{i\omega r_{ij}}{c}\right) \frac{3(\mathbf{n}_{ij} \cdot \mathbf{p}_j) \mathbf{n}_{ij} - \mathbf{p}_j}{r_{ij}^3} + \frac{\omega^2}{c^2} \frac{\mathbf{p}_j - (\mathbf{n}_{ij} \cdot \mathbf{p}_j) \mathbf{n}_{ij}}{r_{ij}} \right] \exp\left(\frac{i\omega r_{ij}}{c}\right). \quad (2.2)$$

Here $\mathbf{E}_{inc,i}$ is the incident field at $i$ th dipole, and $\mathbf{G}(\mathbf{R}_i, \mathbf{R}_j)\mathbf{p}_j$ represent the field radiated by $j$ th dipole at the position of $i$ th dipole; $\mathbf{R}$ is the vector of coordinate of a dipole; $\mathbf{r}_{ij} = \mathbf{R}_i - \mathbf{R}_j$; $\mathbf{n}_{ij} = \mathbf{r}_{ij}/r_{ij}$ is a unit vector parallel to $\mathbf{r}_{ij}$; $c$ is the speed of light. Equation (2.1) can be rewritten as,

$$\frac{1}{\alpha}\mathbf{p}_i - \sum_{j \neq i} \mathbf{G}(\mathbf{R}_i, \mathbf{R}_j) \mathbf{p}_j = \mathbf{E}_{inc,i}. \tag{2.3}$$

For periodic systems, we apply the periodic boundary conditions,

$$\mathbf{p}_i = \mathbf{p} \exp(i\mathbf{k}_B \cdot \mathbf{R}_i), \tag{2.4}$$

$$\mathbf{E}_{inc,i} = \mathbf{E}_{inc} \exp(i\mathbf{k}_B \cdot \mathbf{R}_i), \tag{2.5}$$

where $\mathbf{k}_B$ is the Bloch wave vector, and the form of external field (Eq.(2.5)) is chosen to excite one particular eigenmode with Bloch vector $\mathbf{k}_B$.

We put Eq.(2.4), (2.5) into Eq. (2.3) and have

$$\mathbf{Mp} = \mathbf{E}_{inc}, \tag{2.6}$$

where $\mathbf{E}_{inc}$ is a vector whose components are the amplitude of external incident wave, $\mathbf{M}$ is a 3 x3 matrix operating on 3D vector $\mathbf{p}$,

$$\mathbf{M} = \frac{1}{\alpha}\mathbf{I} - \boldsymbol{\beta}, \tag{2.7}$$

$\mathbf{I}$ is a unit matrix and $\boldsymbol{\beta}$ is a dyadic of lattice sum of dipolar Green's function

$$\boldsymbol{\beta} = \sum_{\mathbf{R}' \neq 0} \mathbf{G}(\mathbf{0}, \mathbf{R}') \exp(i\mathbf{k}_B \cdot \mathbf{R}'). \tag{2.8}$$

Here we have used the equality $\mathbf{G}(\mathbf{R}_i, \mathbf{R}_j) = \mathbf{G}(\mathbf{0}, \mathbf{R}_j - \mathbf{R}_i)$ and $\mathbf{R}' = \mathbf{R}_j - \mathbf{R}_i$. The dyadic $\boldsymbol{\beta}$ contains the geometrical information.

If the dipoles are arranged in the $x - y$ plane, we can exploit the property of dipolar Green's function to decouple the matrix equation (2.6) to two independent equations

$$M_z p_z = E_{inc,z}, \text{ (Transverse)} \tag{2.9}$$

$$\mathbf{M}_\parallel \mathbf{p}_\parallel = \mathbf{E}_{inc,\parallel}, \text{ (In-plane)} \tag{2.10}$$

where $M_z$ ($\beta_z$) is a scalar and $\mathbf{M}_\parallel$ ($\beta_\parallel$) is a symmetric 2 by 2 matrix

$$M_z = \frac{1}{\alpha} - \beta_z, \tag{2.11}$$

$$\mathbf{M}_\parallel = \frac{1}{\alpha}\mathbf{I} - \beta_\parallel = \begin{pmatrix} \frac{1}{\alpha} - \beta_{\parallel,11} & -\beta_{\parallel,12} \\ -\beta_{\parallel,21} & \frac{1}{\alpha} - \beta_{\parallel,22} \end{pmatrix}. \tag{2.12}$$

For in-plane modes, the $p_x$ and $p_y$, which are two components of $\mathbf{p}_\parallel$ along $x$ and $y$ axis respectively, are generally coupled to each other. We can diagonalize $\mathbf{M}_\parallel$ with an orthogonal transformation $\mathbf{A}$,

$$\mathbf{M'}_\parallel \mathbf{p'}_\parallel = \mathbf{E'}_{inc,\parallel}, \tag{2.13}$$

where

$$\mathbf{M'}_\parallel = \mathbf{A}\mathbf{M}_\parallel \mathbf{A}^T = \begin{pmatrix} M'_{\parallel,11} & 0 \\ 0 & M'_{\parallel,22} \end{pmatrix}, \tag{2.14}$$

$$\mathbf{p'}_\parallel = \mathbf{A}\mathbf{p}_\parallel = \begin{pmatrix} p'_{\parallel,1} \\ p'_{\parallel,2} \end{pmatrix}, \tag{2.15}$$

$$\mathbf{E'}_{inc,\parallel} = \mathbf{A}\mathbf{E}_{inc,\parallel} = \begin{pmatrix} E'_{inc,1} \\ E'_{inc,2} \end{pmatrix}. \tag{2.16}$$

The dipole polarizability for a sphere can be written as

$$\alpha(\omega) = \frac{\varepsilon(\omega) - \varepsilon_m(\omega)}{\varepsilon(\omega) + 2\varepsilon_m(\omega)} a^3, \tag{2.17}$$

where $\varepsilon$ is permittivity of the material of the nano-sphere and the $\varepsilon_m$ is permittivity of the media where spheres are embedded. In this paper we focus on the case in which the spheres are in air where $\varepsilon_m = 1$. After diagonalization, we can find a transversal in-plane mode(TI mode, dipole moment perpendicular to Bloch vector) and a longitudinal in-plane mode(LI mode, dipole moment parallel to Bloch vector), which are decoupled to each other.

To take care of the radiation correction, we write[26]

$$\frac{1}{\alpha} \rightarrow \frac{1}{\alpha} - i\frac{2\omega^3}{3c^3} \tag{2.18}$$

such that

$$\frac{1}{\alpha} = \left(1 + \frac{3}{\varepsilon - 1}\right)\frac{1}{a^3} - i\frac{2\omega^3}{3c^3}. \tag{2.19}$$

**III. QUASI-STATIC LIMIT**

We first consider the quasi-static response for a 2D periodic array of lossless metal nano-spheres. We use the Drude model to describe the permittivity as a function of frequency near the plasma resonance,

$$\varepsilon(\omega) = 1 - \frac{\omega_p^2}{\omega(\omega + i\upsilon)} \tag{3.1}$$

where $\omega_p$ is the plasma frequency.

As all dipoles should be decaying in time, the normal mode frequency $\omega$ of the array should be complex with a negative imaginary part. However, because of the factor $e^{i\omega r_{ij}/c}$, the lattice sum **β** diverges when $\text{Im}(\omega) < 0$.

In the quasi-static limit, the speed of light is taken to be infinity ($c \rightarrow \infty$). In addition, because in a lossless metal $\upsilon = 0$, there is no energy loss and the polarizability has the simple form

$$\frac{1}{\alpha} = \frac{1}{a^3}\left(1 - \frac{\omega^2}{\omega_0^2}\right), \tag{3.2}$$

where $\omega_0 = \omega_p/\sqrt{3}$ is the plasma resonance frequency for the sphere.

In the absence of radiation damping and current dissipation, the quality factor for each eigenmode (transverse or in-plane) is infinity. That means the system has no coupling to the external environment. As a result, to investigate the eigenmodes of the system, we can simply set the incident wave $\mathbf{E}_{inc}$ to be zero. The solutions to dispersion equations (2.9) and (2.13) are then $M_z = M'_{11} = M'_{22} = 0$. The detailed form of dispersion relations for the eigenmodes of 2D periodic array of lossless metal nanospheres in quasi-static limit is then

$$\begin{cases} \omega^2 = \omega_0^2 \left(1 - a^3 \beta_z\right), (Transverse) \\ \omega^2 = \omega_0^2 \left\{1 - \dfrac{a^3}{2}\left[\beta_{\parallel,11} + \beta_{\parallel,22} + \sqrt{\left(\beta_{\parallel,11} - \beta_{\parallel,22}\right)^2 + 4\beta_{\parallel,21}^2}\right]\right\}, (Transversal\ In\text{-}plane) \\ \omega^2 = \omega_0^2 \left\{1 - \dfrac{a^3}{2}\left[\beta_{\parallel,11} + \beta_{\parallel,22} - \sqrt{\left(\beta_{\parallel,11} - \beta_{\parallel,22}\right)^2 + 4\beta_{\parallel,21}^2}\right]\right\}. (Longitudinal\ In\text{-}plane) \end{cases} \quad (3.3)$$

Here in quasi-static limit, the **β** is a dyadic of lattice sum of dipolar Green's function in which the retardation is disregarded,

$$\beta_z = -\sum_{\mathbf{R} \neq \mathbf{0}} \frac{1}{R^3} \exp(i\mathbf{k_B} \cdot \mathbf{R}), \quad (3.4)$$

$$\boldsymbol{\beta}_{\parallel} \equiv \begin{pmatrix} \beta_{\parallel,11} & \beta_{\parallel,12} \\ \beta_{\parallel,21} & \beta_{\parallel,22} \end{pmatrix} = \sum_{\mathbf{R} \neq \mathbf{0}} \frac{\exp(i\mathbf{k_B} \cdot \mathbf{R})}{R^3} \begin{pmatrix} 3R_x^2/R^2 - 1 & 3R_xR_y/R^2 \\ 3R_xR_y/R^2 & 3R_y^2/R^2 - 1 \end{pmatrix}. \quad (3.5)$$

As **β** can be expressed in $1/d^3$ times a dimensionless number that is independent of $d$, we can conclude from Eq. (3.3) that the dispersion relations have band-widths proportional to $a^3/d^3$ which corresponds to the coupling strength, just the same as that case for 1D chain[11]. Because of such kind of coupling coefficient, we can enhance the coupling effect among dipoles by increasing the radius and decreasing the lattice spacing. However, we should also keep in mind that in order to guarantee the accuracy of the dipolar approximation, the radius $a$ must be small enough and the lattice spacing $d$ is required not to be less than $3a$; otherwise high-order Mie resonance beyond the dipolar resonance will emerge[27]. To obtain results comparable with previous work, we set the radius of a sphere $a = 25\text{nm}$ and the lattice spacing $d = 75\text{nm}$. The result for an infinite square lattice of silver spheres is plotted in Figure 1.

From Figure 1 we can see that at the Γ point, the center of Brillouin Zone (BZ), the two in-plane modes are degenerate at about $0.91\omega_0$ which is close to that of longitudinal mode for 1D chain, while the resonant frequency for transverse mode (T mode) is about $1.15\omega_0$ and higher than that for 1D chain. The behaviors of the three modes at different symmetric points of the first BZ can be explained by a simple restoring-force model. Within the quasi-static limit, the dipolar Green's function contains only the short-range interaction varying with distance in the form of $1/R^3$, it is thus reasonable to consider the effect of only those nearest and second-nearest dipoles to the reference dipole. In order to understand qualitatively the dispersion relations at symmetric points of first BZ, we plot in Figure 2 the relative phase of dipole moments distributed near the reference particle for different symmetric points of first BZ of square lattice.

For a single particle, the plasmonic resonant frequency ($\omega_0$) of the induced charges is provided by a self-induced restoring force that is due to the Coulomb attraction between the positive cores and the displaced negative electron charges. When there are many particles close together, the net surface charges on the other particles will provide additional Coulomb forces acting on the negative charge on the first particle, thus change the net restoring force. An increase (decrease) in restoring force will increase (decrease) the resonant frequency. To investigate these interaction forces, we summarize the orientations of the dipole moments for different symmetry points in Fig. 2.

For T mode, the dipole moments are perpendicular to the plane of the particle array. If the dipole moments of two neighboring particles are in-phase (anti-phase), the net surface charges on the other particle will add a force on the plasma of the first particle in a direction that is opposite to (the same as) the displacement of plasma, thus increase (decrease) the restoring force. At Γ point, all dipoles are in-phase. As a result, the resonant frequency, $\omega_T(\Gamma)$, is higher than $\omega_0$. At X point, we can see that the electric fields from the four nearest dipoles cancel each other so that the forces provided by the four anti-phase second-nearest dipoles become important. As a result the resonant frequency, $\omega_T(X)$, is slightly lower than $\omega_0$. At M point, the four nearest dipoles are all anti-phase with the reference dipole. Therefore, the overall reduction in the restoring force is more than that for X point, leading to a lower resonant frequency than that for X point, namely $\omega_T(M) < \omega_T(X)$.

For in-plane modes, the dipole moments are confined along the plane of particle array. The interaction forces, in this case, can be classified into two kinds. We call the force from a parallel dipole that is positioned along (perpendicular to) the polar direction the longitudinal (transverse) interaction force. By inspecting the field generated by a dipole at a fixed short distance, one can easily see that the longitudinal force is larger than the transverse force. For the TI mode at Γ point, two nearest particles provide longitudinal forces that decrease the restoring force while the other two nearest particles provide transverse forces that increase the restoring force. Therefore, the final result is a reduction in restoring force, which results in a lower resonant frequency, namely $\omega_{I1}(\Gamma) < \omega_0$. Since TI mode and LI mode has the same dipole arrangements (except a 90° rotation) at Γ point, they are degenerate and thus we have $\omega_{I1}(\Gamma) = \omega_{I2}(\Gamma) < \omega_0$. By similar analysis, one can see that the resonant frequency of the TI (LI) mode at the X point is higher (lower) than that at the Γ point. Therefore, we have $\omega_{I1}(X) < \omega_0$ and $\omega_{I2}(X) > \omega_0$. The two in-plane modes are also degenerate at M point, and has a higher resonant frequency than $\omega_0$.

Due to the geometrical symmetry, the slopes of the dispersions at zone center should be exactly zero. However, Fig. 2 shows that the $v_g$ at zone-center appears to have a finite slope near the zone center. To reveal the physics behind this strange behavior, a more detailed and careful inspection of functions near the zone-center is required. This can be done by taking a straightforward differentiation of Eq. (3.3) at Γ point. We will show that an integral method for full-dynamic case followed by taking the limit $c \to \infty$ in $k_\omega = \omega/c$ will be helpful for explaining the above issue. We will return to this issue in next section.

## IV. FULL DYNAMIC RESPONSE
**a.** *Effective Polarizability*

In this section we will discuss full dynamic response of the 2D infinite array of metal nanospheres. We consider the retardation effect and the intrinsic loss of metal spheres. We solve this problem as a driven system in which the plasmonic spheres respond to an external driving field with a driving frequency that is a real number to investigate the steady-state of a quasi-mode (resonant mode). As we stick with the real axis in frequency, the divergence problem we mentioned earlier for the quasi-static case will not occur. With incident driving field and a finite current dissipation $\upsilon$, the extinction cross section of the whole system is non-zero. From Eqs. (2.9), (2.13) and (2.14), we have

$$p_z = \alpha_{eff,z} E_{inc,z}, \qquad (4.1)$$

$$p'_{\|,1} = \alpha_{eff,1} E'_{inc,1}, \qquad (4.2)$$

$$p'_{\|,2} = \alpha_{eff,2} E'_{inc,2}, \qquad (4.3)$$

where

$$\alpha_{eff,z} = \frac{1}{M_z} \qquad (4.4)$$

$$\alpha_{eff,1} = \frac{1}{M'_{\parallel,11}},. \quad (4.5)$$

$$\alpha_{eff,2} = \frac{1}{M'_{\parallel,22}}. \quad (4.6)$$

Here the $\alpha_{eff}$ is an effective polarizability that represents the effective response of a reference particle to the external incident field. This effective polarizability is contributed by not only the intrinsic polarizability of a metal nano-sphere but also the scattering from other nanospheres located at periodic lattice sites. Hence, we have considered both single-particle properties and inter-particle coupling in which geometrical information and periodic boundary condition are included. From the derivation of $\alpha_{eff}$ we note that in an infinite lattice (with one particle per unit cell) each particle has the same effective polarizability. Once the effective polarizability is known, the extinction cross section for each sphere is readily obtained from the optical theorem[28],

$$C_{ext} = \frac{4\pi\omega}{c}\text{Im}(\alpha_{eff}). \quad (4.7)$$

As a function of the driving frequency, the imaginary part of $\alpha_{eff}$ will exhibit peak(s)[15], which are in good agreement with the full dynamic dispersion relations obtained by finite-chain solution[11]. In addition, the mode quality-factors can also be obtained from the full-width at half maximum (FWHM) of the peak. The advantage of this method is that it enables us to simultaneously see the full-dynamic dispersion relations and the quality-factors for quasi-modes.

We can easily obtain the resonant frequencies at a fixed spatial wave-vector $k_S$, by seeking peak(s) of $\text{Im}(\alpha_{eff})$ as a function of $\omega$, as well as the dispersion relations by showing loci of resonant frequencies for varied $k_S$ in the first BZ. Second, the bandwidth of the resonant peak(s) of $\text{Im}(\alpha_{eff})$ can be related to the mode quality-factor by $Q = \omega_r / \Delta\omega$, where $Q$ is the quality factor, $\omega_r$ is the resonant angular frequency and $\Delta\omega$ is the bandwidth (FWHM) of the quasi-mode. As a result, by plotting the map of $\text{Im}(\alpha_{eff})$ as a function of spatial wave-vector $k_S$ and incident frequency $\omega$, we can observe both the dispersion relations and the mode qualities.

**b.** *Lattice Sum of Green's Function*

Evaluating $\alpha_{eff}$ requires the evaluating the lattice sum of Green's function $S$. In the dynamic dipolar Green's function, a term proportional to $e^{i\mathbf{k_B}\cdot\mathbf{R}}e^{i\omega R/c}/R$ represents the long-range interaction, while the other terms represent the short-range ($\sim 1/R^3$) and intermediate-range ($\sim 1/R^2$) interaction. The lattice sums of short-range term and intermediate-range terms have good convergence and can be calculated numerically with a satisfactory accuracy, as long as the spatial lattice used in the calculation is large enough. However, the lattice sum of the long-range term does not converge for real $\omega$. In the literature, several methods have been developed to deal with such kind of lattice sum. A special method for summing the long-range term with the help of imaginary dipole array has been discussed by Belov *et al.* in Ref. 29. If the dipole array is dense, namely the dipole spacing is far less than the incident wavelength, an integral method can be applied to evaluate that lattice sum[19,29].

More rigorous and general methods are related to the application of Poisson's formula and the summation on reciprocal lattice. For summing the long-range term on infinite and completed 2D spatial lattice, the Ewald's method can be applied to split the term into two parts, with one fast-converging on the spatial lattice and the other on reciprocal lattice[30]. In our case, although the spatial lattice where the summation will be done is not completed because of the exclusion of origin, we can still re-write the summation as a complete one accompanied by a self-radiated diverging term subtracted (See Eq.(A2) in Appendix A). Actually a straightforward application of Ewald's method to the first complete series in Eq.(A2) is proven efficient in generating numerically accurate and quickly-converging results, even when $z \to 0_+$. However, for summing such kind of long-range term on uncompleted lattice, Simovski et al. have developed an interesting series[18] which will be adopted below. The method can help identify more clearly the mathematical contributions from different physical parameters.

While the form of the series in Ref. [18] was for rectangular lattice only, the idea is valid for arbitrary 2D periodic arrays with one particle in one unit cell. The original paper of Simovski et al.[18] contains some typos, and for that reason, we re-derived the fast-converging series following the work in Ref. [18] in Appendix A. Our result is more general and is valid for any 2D simple lattice. The result is

$$S(k_\omega, \mathbf{k_B}) \equiv \sum_{\mathbf{R} \neq 0} \frac{e^{i(k_\omega R + \mathbf{k_B} \cdot \mathbf{R})}}{R}$$
$$= D + \frac{2\pi i}{\Omega} \frac{1}{k_{z,0}} - ik_\omega + \frac{2\pi}{\Omega} \sum_{\mathbf{G} \neq 0} \left( \frac{1}{k_{z,\mathbf{G}}} - \frac{1}{G} \right), \quad (4.8)$$

where D is a constant for a specific lattice structure, $\mathbf{G}$'s are 2D reciprocal lattice vectors; $\Omega$ is the area of one unit cell in real-space lattice; $k_\omega = \omega/c$ is wave number of incident wave; $k_{z,0} = \sqrt{k_\omega^2 - |\mathbf{k_B}|^2}$, $k_{z,\mathbf{G}} = \sqrt{|\mathbf{k_B} + \mathbf{G}|^2 - k_\omega^2}$ and $\text{Im}(k_{z,0}) \geq 0$ and $\text{Im}(k_{z,\mathbf{G}}) \geq 0$.

Given a fixed $k_\omega$, as $\mathbf{k_B}$ increases and cross the light-line, the second term $\frac{2\pi i}{\Omega} \frac{1}{k_{z,0}}$ in Eq.(4.8) experiences a divergence and goes from purely imaginary to purely real. It hence implies that the dispersions could also experience a singularity, which we will show below, at the light-line.

The interesting quantity $D$ is found to possess definite physical meaning. As we can see from its definition (see Appendix A), $D$ is represents a *geometrical* effect in electrostatic limit as $k_\omega$ tends to zero, and it only depends on the lattice structure rather than any wave nature of the summation. On the other hand, both the second term and third term in Eq. (4.8) depend on the wave number $k_\omega$ and wave vector $\mathbf{k_B}$ so that they mainly represent the *wave* nature of the summation. Last, the series of correction terms are evaluated over reciprocal lattice excluding origin so that it depends on both the geometrical effect and the wave nature. However, the correction series is much smaller than other terms. As a result, the series is dominated by the first three terms and is very fast-converging.

When we evaluate $D$, we can replace the series in Eq.(A8) in Appendix A with a two-dimensional integral in the reciprocal space,

$$\Omega_G \sum_G{'} \to \int_0^{2\pi} d\theta \int_{g_{\min}}^{\infty} g\, dg, \quad (4.9)$$

where $\Omega_G = \frac{4\pi^2}{\Omega}$ is the area of unit cell of reciprocal lattice and we have assumed that the replacement is rigorous as long as the integral is taken outside a finite circle centered at origin and with radius $g_{\min}$. Therefore,

$$D = \lim_{z \to 0_+} \left( \frac{\Omega_G}{2\pi} \sum_G{'} \frac{e^{-zG}}{G} - \frac{1}{z} \right) = \lim_{z \to 0_+} \left( \frac{1}{2\pi} \int_0^{2\pi} d\theta \int_{g_{\min}}^{\infty} g\, dg \frac{e^{-zg}}{g} - \frac{1}{z} \right)$$
$$= \lim_{z \to 0_+} \left( \int_{g_{\min}}^{\infty} e^{-zg} dg - \frac{1}{z} \right) = \lim_{z \to 0_+} \left( \frac{e^{-zg_{\min}}}{z} - \frac{1}{z} \right) = -g_{\min}. \quad (4.10)$$

In Eq. (4.10), we have shown that $D$ must be a negative real number with a magnitude $g_{min}$ which is the radius of a finite circle, and we have assumed that the infinite integral taken out of that circle is exactly equivalent to the infinite lattice sum in reciprocal space. We introduce a parameter $U$ sufficiently large and let the summation in Eq. (4.10) be numerically calculated in the region $0 < G \leq U/z$. We then define a new quantity

$$L \equiv \frac{\Omega_G}{2\pi} \sum_{0 < G \leq U/z}' \frac{e^{-zG}}{G} \xrightarrow{z \to 0_+} L = \int_{g_{min}}^{U/z} e^{-zg} dg = \frac{e^{-zg_{min}} - e^{-U}}{z}, \quad (4.11)$$

Here we again employ our assumption that an infinite lattice summation can be rigorously replaced by a 2D integral with $g_{min}$ as inner bound and infinity as outer bound. Hence we have two methods of evaluating $D$ by approaching to the zero-point limit of two different function,

$$D = -g_{min} = \lim_{z \to 0_+} f_1(z) = \lim_{z \to 0_+} \frac{\ln(zL + e^{-U})}{z}, \quad (4.12)$$

and

$$D = \lim_{z \to 0_+} f_2(z) = \lim_{z \to 0_+} \left( L + \frac{e^{-U} - 1}{z} \right). \quad (4.13)$$

In Figure 3 we plot the $f_1(z)$ and $f_2(z)$ for square lattice in small $z$ region and find excellent linear convergence for the two functions. Extrapolations at $z = 0$ are done for the two lines and the results coincide with each other to the order of $10^{-4}$. Then we take the averaged number of the two extrapolated results to be final result of $D$. For square lattice $D = -3.9002/d$, in good agreement with the result obtained by Simovski *et al.*[31]; with similar approach, we obtain $D = -4.2133/d$ for triangular lattice. The parameter $U$ can of course be increased to obtain arbitrary precision, but the current results are good enough for our purpose.

**c.** *Results and discussion*

For the three modes propagating on an infinite 2D nano-sphere square lattice, we plot in Fig.4 the dimensionless $\text{Im}(\alpha_{eff})/a^3$ in a color-map as a function of the incident frequency $\omega$ and wave vector $k_B$ in ΓX direction within the first Brillouin Zone. The short-range and intermediate-range terms in dipolar Green's function are summed over $3000 \times 3000$ grids, while the long-range terms are evaluated by Eq.(4.8), with the correction terms summed in the region $|\mathbf{G}| \leq 320\pi/d$. In the numerical calculation we use a Drude model fitted from experimental data[32],

$$\varepsilon(\omega) = \varepsilon_a - \frac{(\varepsilon_b - \varepsilon_a)\omega_{pl}^2}{\omega(\omega + i\upsilon)} \quad (4.14)$$

with parameters $\varepsilon_a = 5.45$, $\varepsilon_b = 6.18$, $\omega_{pl} = 11.34 \, \text{eV}$ (single-sphere resonant frequency $\omega_0 = 3.57 \, \text{eV}$), $\upsilon = 0.05 \, \text{eV}$ to obtain results for silver nano-spheres. All the spheres have same radius of 25nm and a center-to-center spacing of 75nm, the same as our quasi-static calculation and some previous work by others.

From Figure 4, we can observe the following phenomena of the three modes: (1) Inside the light-cone, the T-mode is rather sharp near the zone center while the two in-plane modes are fuzzy; (2) the bands of T mode and TI mode drops suddenly and significantly when they meet the light-line, while for LI mode the band drops down continuously with a small negative $v_g$ before it meets light-line; (3) outside the light-cone, all three modes are well-defined, with relatively high quality-factors.

These observations can be explained as follows. First, all three modes have radiation damping within the light-cone. For a single dipole, the far-field time-averaged Poynting vector $\bar{\rho}$ radiated by this dipole has an angular distribution in the form $\bar{\rho} \propto \sin^2\theta$ [26], where $\theta$ is the angle of Poynting vector against the dipole moment (see Figure 5). For T-mode (in-plane modes), the energy radiated in (out of) the plane is strong. Therefore, inside the light-cone the two in-plane modes are very leaky, while the T-mode has relatively good quality because the energy is kept in the plane. Second, the sharp drop from the zone center towards the light-line has been observed for the T mode of 1D chains [8,11,14,15] and is the result of the strong coupling of the plasmonic mode with free photons in vacuum. We see the same phenomena here in 2D. The T mode and TI mode in 2D arrays are developed from the two degenerate transverse modes of an infinite 1D chain, while the other in-plane LI mode originates from the longitudinal mode of the 1D chain. That is why we call the in-plane modes "transverse /longitudinal in-plane mode (TI/LI mode)". The sudden dips in dispersions for both T mode and TI mode indicate a strong coupling with free photons. With dipole moments orthogonal to the propagating wave vector (and hence parallel to the E-field), these two modes can be directly excited by free photons. In contrast, the LI mode has no such coupling with free photons as the dipole moment is parallel to wave vector. In 1D silver-sphere chain, the longitudinal mode joints smoothly from inside the light-cone to the guided modes outside the light-cone. However, the LI mode of 2D silver-sphere array shows a pseudo coupling with light-line at which the band's slope ($v_g$) shows a negative-to-positive transition. We will show below that the $v_g$ inside the light-cone for the LI mode of this specific 2D array is inherently-negative), and thus the LI mode has to bend downward in frequency before it meets the light-line and has to turn upward when it becomes a guided mode outside the light-cone. Third, outside the light-cone, the phase velocities of all three modes are larger than light-speed $c$, the field must have an exponential decay in the direction perpendicular to the plane and no energy can be radiated. Thus, the guided mode qualities for all three modes are all determined by absorption loss which is dictated by the $\upsilon$ in Drude model.

Because the imaginary part of $M$ is small, the peak of $\text{Im}(\alpha_{eff})$ coincides with the contour $\text{Re}(M) = 0$. As we discussed before, the geometrical constant $D$ extracted out of the lattice sum of long-range interaction is included in $\text{Re}(M)$, therefore it will affect the resonant frequencies. For example, the magnitude of $D$ in triangular lattice is larger than that in square lattice, resulting in a small up-shift of the resonant frequency for transverse mode and a down-shift for in-plane modes in Brillouin zone center. The larger magnitude of $D$ correspond to a more close-packed real-space lattice, and thus the resonant frequency(ies) of transverse (in-plane) mode(s) will be higher (lower) in a more close-packed lattice when all the particles are in-phase.

For the square lattice, the dispersion along ΓM shows similar behavior as in the ΓX direction: Inside the light-cone the transverse mode has good qualities while the two in-plane modes are hardly observable; the T mode and TI mode are strongly coupled with free photons while the LI mode has not real but pseudo coupling to free photons; last. Outside the light-cone, the guided qualities are determined by the absorption loss. We extract the peak positions of $\text{Im}(\alpha_{eff})$ for the three modes along three symmetric directions in first BZ and plot them in Figure 6.

We can see that inside the light-cone the dispersions (as defined by peaks of $\text{Im}(\alpha_{eff})$) are almost isotropic, so that the frequencies only depend on the magnitude of the wave vector $\mathbf{k}_B$, very weakly on its direction. All the three modes have very flat bands inside the light-cone, indicating very small $v_g$. To understand the underlying physics, it is helpful to seek approimate close-form solutions. As the sphere-spacing 75nm is much smaller than the interested wavelengths, which is near or larger than the single-sphere resonant wavelength 348 nm, the whole 2D dipole array can be treated as a dense array so that the lattice sum can be approximated by an integral, and we can therefore obtain full-analytical expressions of the lattice sum of Green's function[19]. We derive the integral form of the lattice sums of dipolar Green's function in Eq. (2.8) for the three modes following the work in Ref.19 (see Appendix B). The results are

$$\beta_T = \frac{2\pi}{\Omega}(I_2 - I_1), \text{(Transverse mode)} \quad (4.15)$$

$$\beta_{T,I} = \frac{2\pi}{\Omega}(I_2 + I_3), \text{(Transversal In-plane mode)} \quad (4.16)$$

$$\beta_{L,I} = \frac{2\pi}{\Omega}(I_1 - I_3), \text{(Longitudinal In-plane mode)} \quad (4.17)$$

where the $I_1$, $I_2$ and $I_3$ are the results of integrals and are slow-varying with wave vector $k_B$ (see Appendix B).

With these results, we can now explain the apparent finite-slope for quasi-static dispersion near the zone center in the previous section. Let us start from a full dynamic situation with a finite $k_\omega = \omega/c$, and we replace the 2D lattice-sums with the integrals whose details are showed in Appendix B. The integral methods should be a very approximation near the zone-center, because the wave-length is much greater than the lattice spacing. Therefore Eq. (3.3) can be re-written as $\omega_\sigma^2 = \omega_0^2(1 - a^3\beta_\sigma)$ where $\sigma$ denotes the mode-type and $\beta_\sigma$ is just the same as in Eq. (4.15)-(4.17). Next, we take a differentiation of $I_\sigma$ with respect to $k_B$, and then take the limit $k_\omega$ to zero (quasi-static), making the derivatives near zone-center for $I_2$ and $I_3$ approach to zero and $O(k_B)$ respectively. For the $I_1$, the result will be determined by the competition between $k_\omega$ and $k_B$. Given that the quasi-static solutions only have physical meaning when $k_\omega \to 0$, if $k_\omega \gg k_B$, which means $k_B = 0$, the $v_g$ will be exactly zero at $\Gamma$ point; otherwise, $k_B \gg k_\omega \to 0$, then we have

$$\left.\frac{\partial I_1}{\partial k_B}\right|_{k_\omega \to 0} = -1 + O(k_B). \quad (4.18)$$

Thus the from Eq. (4.15)-(4.17) the $v_g$ of different modes near zone center shown in Figure 1 can be explained. All modes must have $v_g$ exactly zero at $\Gamma$ point. In the vicinity of the zone-center, the dispersion for TI mode is parabolic, while the slopes for T mode and LI mode have opposite signs and finite magnitudes.

Next, we continue to discuss the dispersion behavior for full-dynamic cases. We now have full-analytical form of $\text{Im}(\alpha_{eff})$ as $\text{Im}(\alpha_{eff,\sigma}) = -\text{Im}(M_\sigma)/|M_\sigma|^2$. And we must stress again that because the imaginary part of $M$ is much smaller than the real part, the peak of $\text{Im}(\alpha_{eff})$ coincides with the contour $\text{Re}(M) = 0$. Then we plot the $\text{Re}(M) = 0$ contours for the three modes in Figure 7, where M can be analytically evaluated due to the integral method. Figure 7a and 7b correspond to the Drude model of Eq.(3.1) and (4.14) respectively. From Appendix B we know that only $I_2$ has $i/\sqrt{k_\omega^2 - k_B^2}$ divergence. Actually $2\pi I_2/\Omega$ is just integral form of $k_\omega^2 S$, which stands for transversal long-range dipole interaction. From Eq. (4.15-17), we can notice that $I_2$ exists for T and TI, but not for LI. So we conclude that it is the transversal long-range dipole interaction that account for the light-line coupling With the same Drude model, the dispersion found by integral method in Fig.7b are in very good agreement with the dispersion relations shown in Figure 6, particularly for small $k_B$, where the wavelength of the Bloch wave is large and the phase distribution of dipole moments is quite slow-varying. Moreover, the small pseudo coupling between LI mode and light-line is now clearly seen. In addition, the flat bands near center of Brillouin zone are reproduced by the integral methods.

However, for estimating the $v_g$, the more concise form of Drude model Eq. (3.1) is employed in the following discussion. For an implicit dispersion relation $f(\omega, k_B) = 0$ where $f = \text{Re}(M)$, the $v_g$ is given by

$$\frac{d\omega}{dk_B} = -\frac{\partial f / \partial k_B}{\partial f / \partial \omega}. \tag{4.19}$$

Noting that

$$\frac{\partial \text{Re}(M_\sigma)}{\partial \omega} \approx \frac{\partial \text{Re}(\frac{1}{\alpha})}{\partial \omega} = -\frac{2\omega}{a^3 \omega_0^2}, \tag{4.20}$$

$$\frac{\partial \text{Re}(M_\sigma)}{\partial k_B} = -\frac{\partial \text{Re}(\beta_\sigma)}{\partial k_B}, \tag{4.21}$$

we put Eq.(2.19), (3.1), (4.20) and (4.21) into (4.19), and have

$$\frac{d\omega_\sigma}{dk_B} = -\frac{\partial \text{Re}(M_\sigma)}{\partial k_B} / \frac{\partial \text{Re}(M_\sigma)}{\partial \omega} \approx -\frac{a^3 \omega_0^2}{2\omega_\sigma} \frac{\partial \text{Re}[\beta_\sigma]}{\partial k_B} \equiv \frac{\gamma_\sigma k_B \omega_0^2}{\omega_\sigma}, \tag{4.22}$$

where the parameter $\gamma_\sigma$ depends on the mode σ. For the T mode, we have (see Appendix B, and in the following $R$ denotes $R_{\min}$ for short)

$$\text{Re}[\beta_T(k_\omega, k_B)] \approx \frac{2\pi}{\Omega} \left\{ \frac{3[\cos(k_\omega R) - \sin(k_\omega R)]}{4R} (k_B R)^2 - \frac{\cos(k_\omega R) + k_\omega R \sin(k_\omega R)}{R} \right\}, \tag{4.23}$$

$$\gamma_T = 12[\sin(k_\omega R) - \cos(k_\omega R)]\gamma_0; \tag{4.24}$$

and for the TI mode,

$$\text{Re}[\beta_{TI}(k_\omega, k_B)] \approx \frac{2\pi}{\Omega} \left\{ \frac{7\cos(k_\omega R) - 5\sin(k_\omega R)}{16R} (k_B R)^2 + \frac{\cos(k_\omega R) - k_\omega R \sin(k_\omega R)}{2R} \right\}, \tag{4.25}$$

$$\gamma_{TI} = [5\sin(k_\omega R) - 7\cos(k_\omega R)]\gamma_0; \tag{4.26}$$

and the for LI mode,

$$\text{Re}[\beta_{LI}(k_\omega, k_B)] \approx \frac{2\pi}{\Omega} \left[ \frac{-3\cos(k_\omega R) + 9\sin(k_\omega R)}{16R} (k_B R)^2 + \frac{\cos(k_\omega R) - k_\omega R \sin(k_\omega R)}{2R} \right], \tag{4.27}$$

$$\gamma_{LI} = [-9\sin(k_\omega R) + 3\cos(k_\omega R)]\gamma_0, \tag{4.28}$$

where $\gamma_0 = \frac{\pi a^3 R}{8\Omega}$.

Thus the dispersion relations for small wave vector in the light-cone are approximately

$$\omega_\sigma^2(k_B) \approx \omega_\sigma^2(0) + \gamma_\sigma \omega_0^2 k_B^2, \tag{4.29}$$

Note that $a/d = 1/3$, $R = d/1.438$ and $\Omega = d^2$ for square lattice, the magnitude of $\gamma_0 / d^2$ is as small as 0.01; moreover, by putting $k_\omega R \simeq 1$ (for silver nano-spheres in square-lattice) into Eq. (4.24), (4.26), (4.28), we obtain $\gamma_T \simeq 3.61\gamma_0, \gamma_{TI} \simeq 0.43\gamma_0, \gamma_{LI} \simeq -5.95\gamma_0$. As a result, the dispersion relations are very flat for small $k_B$ in the light-cone, with the magnitudes of $v_g$

$$|v_{g,\sigma}| = \left| \frac{d\omega_\sigma}{dk_B} \right| \approx \left| \frac{\gamma_\sigma k_B \omega_0^2}{\omega_\sigma} \right| \approx \left| \frac{\gamma_0}{d^2} (k_B d) \left( \frac{\omega_0 d}{c} \right) c \right| \sim |(k_B d) \times 0.01c|. \tag{4.30}$$

In addition, in the light-cone the dispersion relations as given by Eq. (4.29) agrees quite well agree with those obtained by $\text{Re}(M) = 0$ contours (see Figure 7).

With the help of integral method, we have obtained approximate but analytical expressions of the dispersion relations for small wave vectors. In contrast to propagating plasmonic modes of 1D chain[8,11,14,15], the dispersion relations for the quasi-modes of 2D array are much more slow-varying with $k_B$ in the light-cone. Mathematically, the "flat band" phenomena are associated with the slow-varying special functions $J_0, J_1$ in small $k_B$ regime. Also the pseudo light-line coupling for LI mode can now be understood more. Although the down-and-up feature in LI dispersion looks similar to that for coupling with free photons, the mechanisms behind are different: once the dispersions reach the light-line, the inter-particle coupling coefficient of T and TI modes diverges, indicating a strong renormalization, while the LI mode comes to a minimal resonant frequency, which is required by the negative $\gamma_{LI}$ in Eq. (4.28) coming from some special functions particular for 2D systems. Although $\gamma_{LI}$ is related to lattice structure and resonant frequency through $k_\omega R$, within a quite broad range of $k_\omega R$ the $\gamma_{LI}$ is kept negative, hence the negative $v_g$ and pseudo coupling with light-line will also be seen in 2D systems other than square silver-sphere arrays, as long as the integral method is still valid. Also, within this method any lattice-dependent anisotropy is eliminated by the special functions for 2D systems and a small $\gamma_0 / d^2$ is always reached, hence isotropic and flat bands inside light-cone also exist in other 2D dense array with different structure or material. Due to the high quality and nearly-zero $v_g$ for transverse mode, 2D metal nano-spheres array is a good candidate for optical storage, optical cavity, subwavelength imaging[9] near the resonant frequency.

**V. CONCLUSION**

In summary, we have investigated the plasmonic modes on an 2D periodic arrays. The particle arrays are treated in coupled dipole approximation. We solved the problem first in the quasi-static limit and then full-dynamical (exact). In the full-dynamic solution, the dispersion relations corresponding to quasi-modes are obtained through an eigen-decomposition method, which gives an effective polarizability describing the collective response of an infinite system to external driving monochromatic plane wave. The dispersion relations and quality-factors are simultaneously related to the imaginary part of the effective polarizability, which is affected separately by the single-particle properties and inter-particle coupling presented by the lattice sum of dipolar Green's functions. This method allows for the discussion of leaky modes inside the light-cone.

For the quasi-static case, the relative positions of eigen-frequencies of three eigenmodes, particularly at symmetric points in the first BZ, can be explained by a simple restoring-force model. For full-dynamic case, inside the light-cone, transverse mode has a high mode quality while the two in-plane modes show stronger radiation damping. Similar to the case of 1D chain, near the light-line, direct coupling with free photons leads to significant energy loss of transverse mode and transversal in-plane mode. Moreover, inter-particle coupling in a 2D system leads to negative group velocity for longitudinal in-plane mode inside the light-cone. The mode qualities outside the light-cone are dominated by the absorption loss of single-particle. More interestingly, we demonstrate that the isotropic and flat bands observed inside the light-cone can be qualitatively explained by an integral method which gives analytic expressions of dispersion relations. For 2D metal-sphere periodic arrays, we trace the small group velocities of the leaky-modes within light-cone to an intrinsic property of a dense array,.

**Acknowledgement**

We thank Jack Ng, Huanyang Chen and Jun Mei for helpful discussion, and Ting Li for her kindly help on the preparation of Figure 2. This work is supported by a Hong Kong Central Allocation grant HKUST3/06C. Computation resources are supported by Shun Hing Education and Charity Fund.

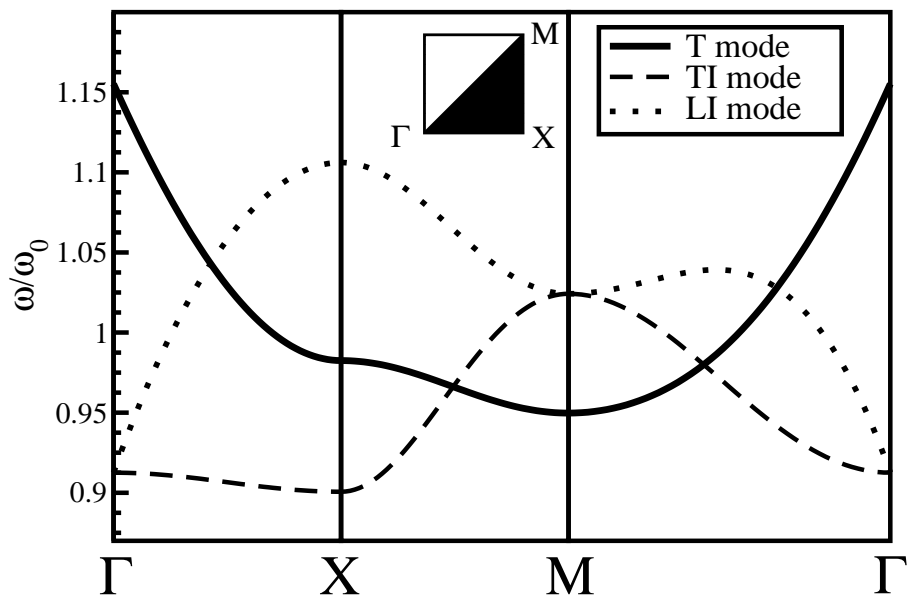

**Figure 1.** Dispersion relations for dipolar transverse mode(solid line), transversal in-plane mode (long dash line) and longitudinal in-plane mode (short dash line) in the quasistatic limit for an infinite square array of lossless 25nm radii spheres spaced by 75nm. The lattice sum is evaluated in a 2000x2000 grid.

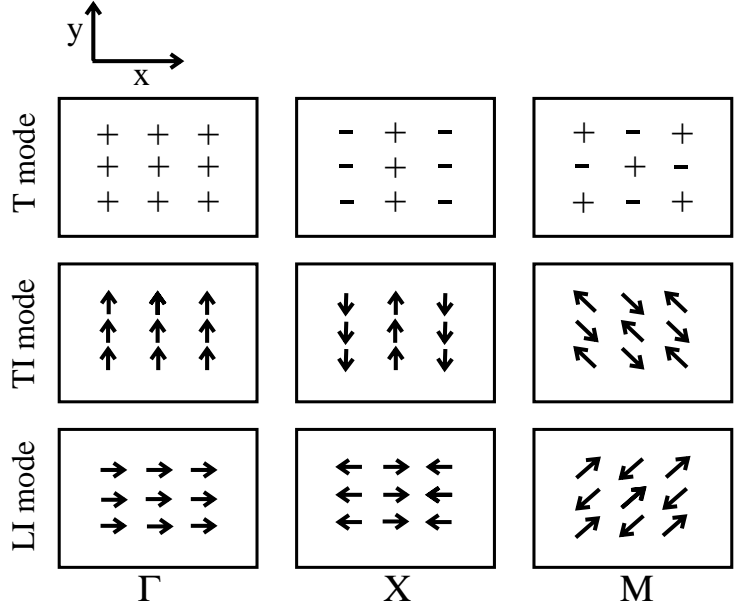

**Figure 2.** The relative phase distribution of dipole moments for different symmetric points of first BZ near the reference particle that is at the center of the 3 by 3 array at $x-y$ plane.

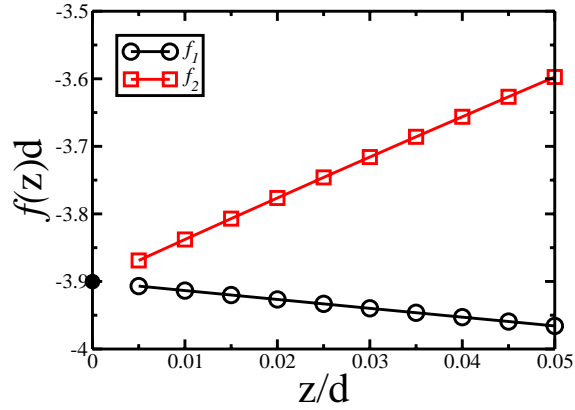

**Figure 3.** (Color online)Plots of $f_1$ (open circle) and $f_2$ (open square) at small $z$ region. The values of $D$, marked by the solid circle at $z=0$ axis, are extrapolated out of $f_1$, $f_2$ and coincident with each other with a difference of order $10^{-4}$. The $U$ is set to be 10 here.

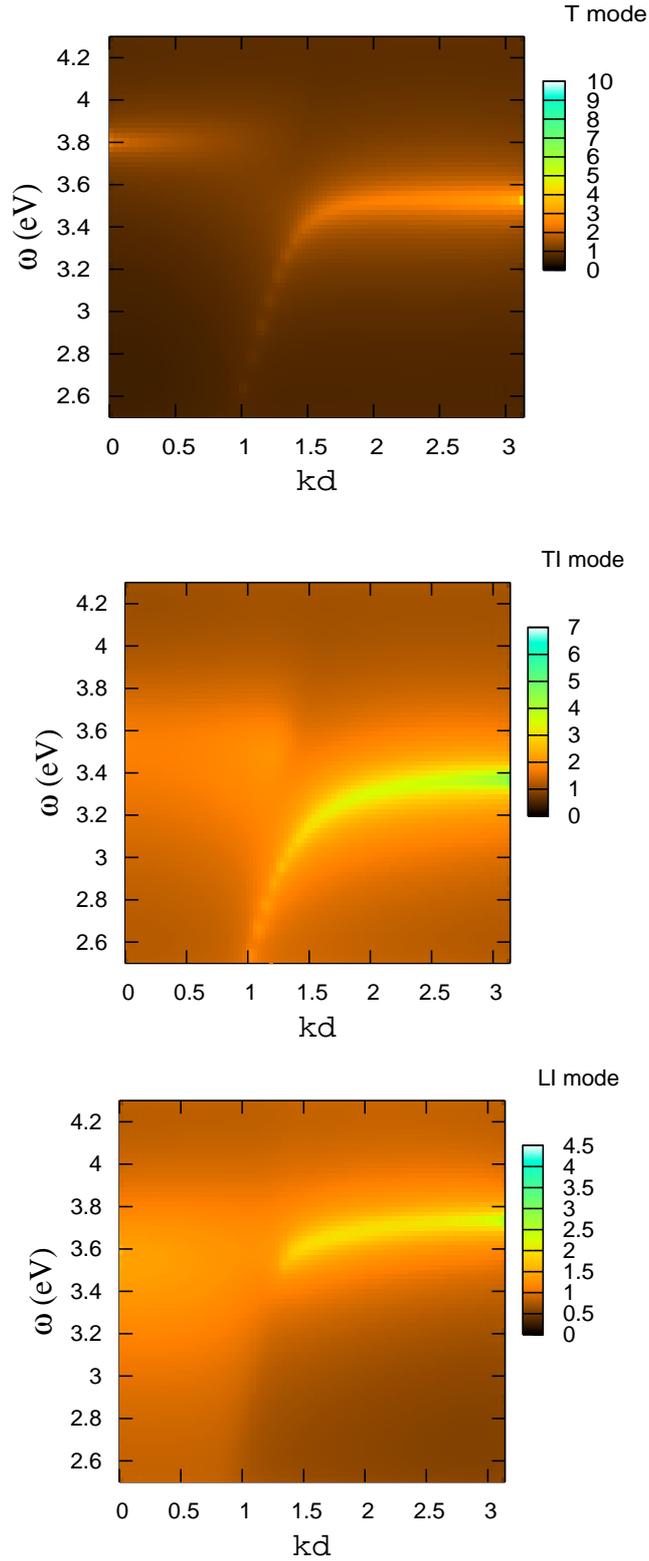

**Figure 4. (Color online) Dispersion diagrams for the quasi-modes of silver nano-spheres array in $\Gamma - X$ section of the first BZ.**

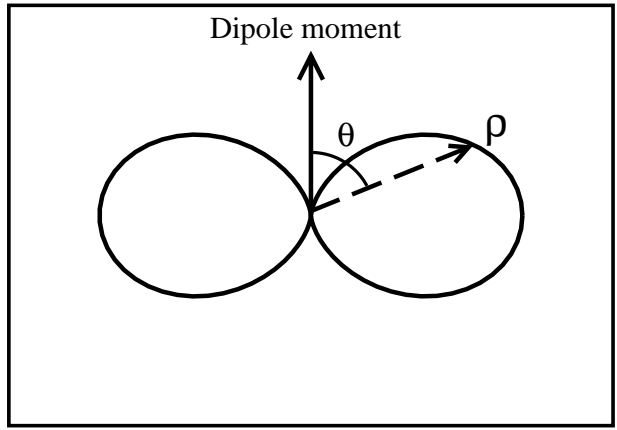

**Figure 5. Illustration for the angular distribution of Poynting vector of the field radiated by a single electric dipole.**

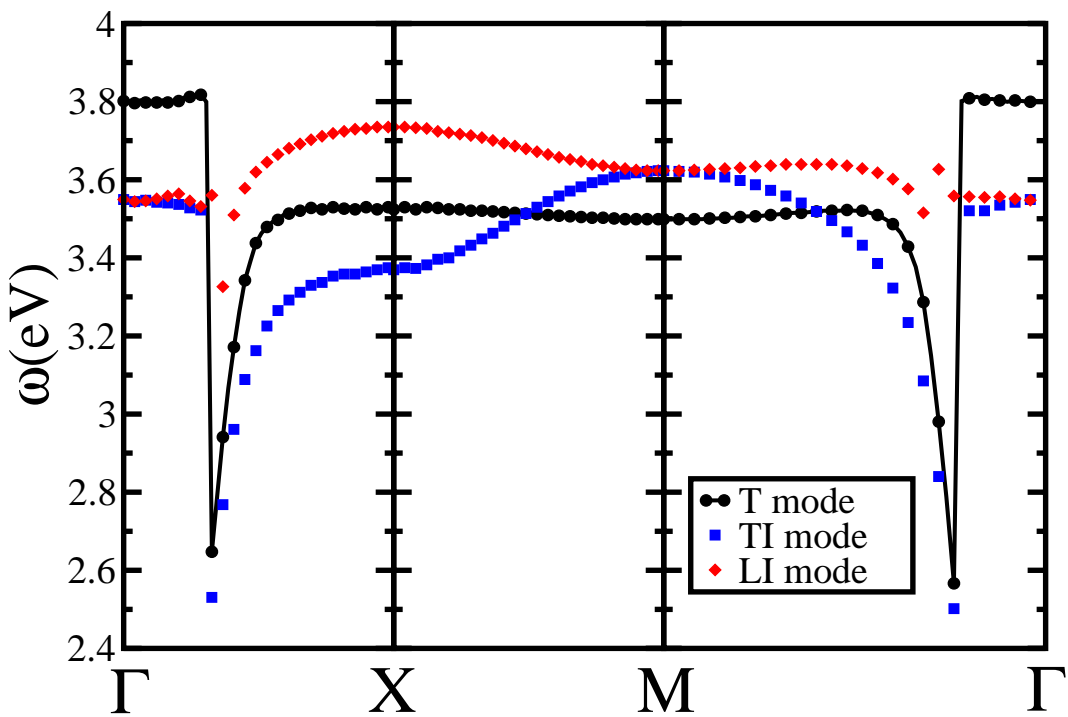

**Figure 6. (Color online) Dispersion relations for the three collective plasmonic modes in first BZ for square lattice. Line with circles represent the transverse mode (T mode); squares represent the transversal in-plane mode (TI mode) and diamonds the longitudinal one (LI mode).**

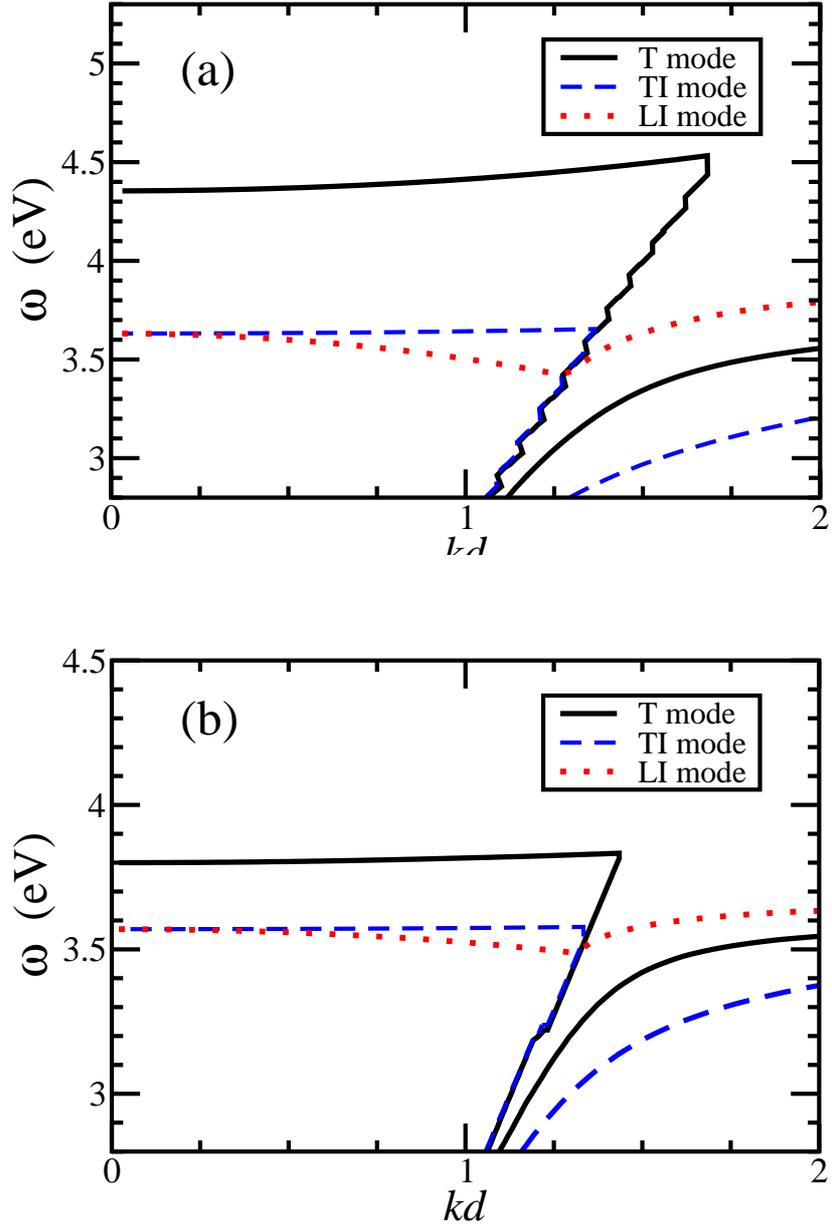

**Figure 7. (Color Online)** $\mathrm{Re}(M) = 0$ **contours for the transverse mode (solid line), transversal in-plane mode (long dash line) and longitudinal in-plane mode (short dash line) in the first Brillouin zone, with Drude Model of (a) Eq. (3.1);(b) Eq.(4.14).**

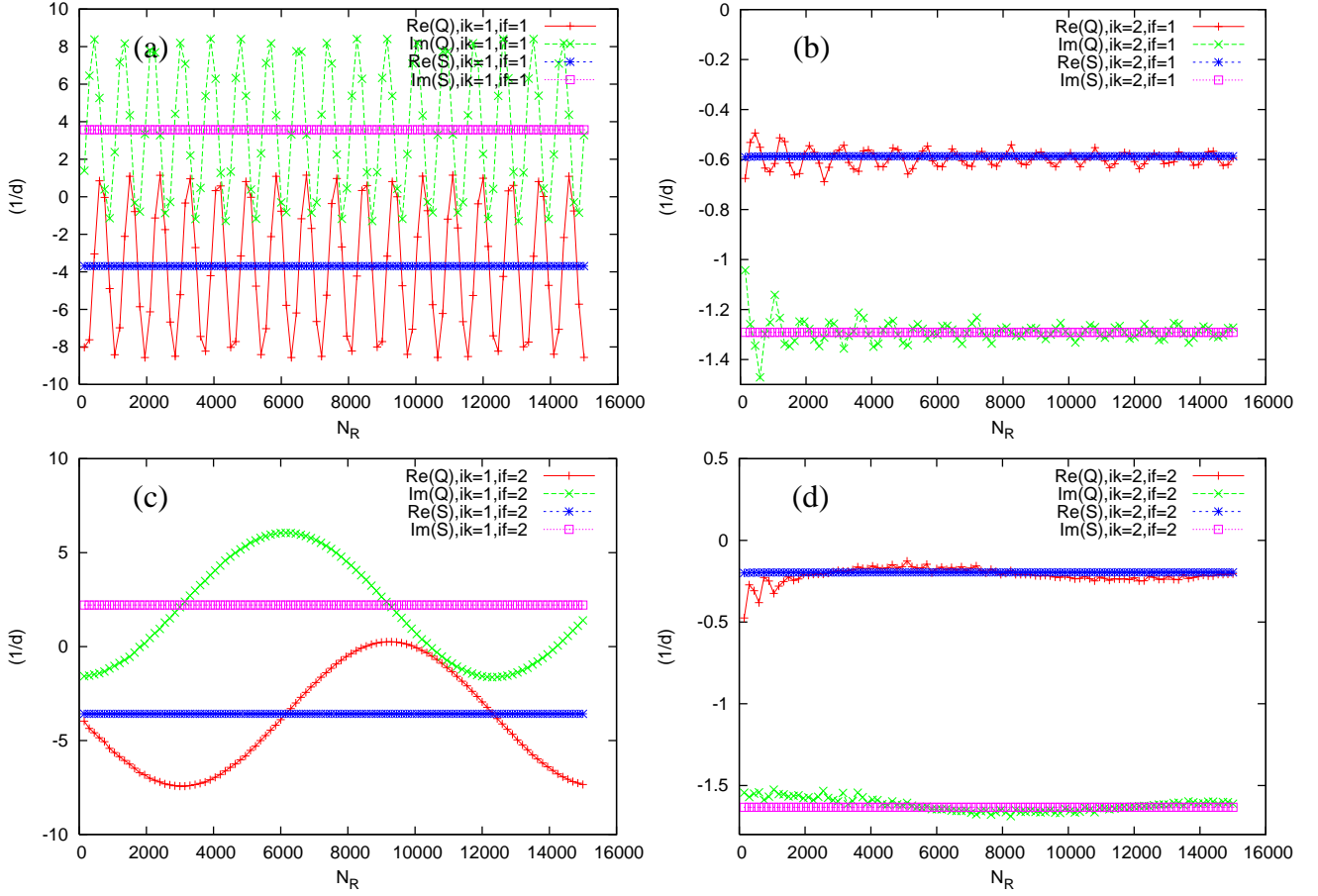

**Figure 8.** (Color online) Examples of the convergence of the Simovski's series (S) compared with the original real-lattice series (Q). The circle within which the series is evaluated has a radius of $N_R d$ ($2\pi N_R / d$) for original (Simovski) series. Lines with "+" ($\times$) represent the real (imaginary) part of the original sum, and lines with stars (open squares) represent the real (imaginary) part of the Simovski's series. The denotation "if"=1,2 are for $\omega < \omega_0$, $\omega > \omega_0$, and "ik"=1,2 are for $k_B < \frac{\omega}{c}$, $k_B > \frac{\omega}{c}$ respectively. **(a)** $\omega < \omega_0$ and $k_B < \frac{\omega}{c}$; **(b)** $\omega < \omega_0$ and $k_B > \frac{\omega}{c}$; **(c)** $\omega > \omega_0$ and $k_B < \frac{\omega}{c}$; **(d)** $\omega > \omega_0$ and $k_B > \frac{\omega}{c}$.

Appendix A.

Let

$$Q(\mathbf{r}, k_\omega, \mathbf{k}_B) \equiv \sum_{\mathbf{R}} e^{i\mathbf{k}_B \cdot \mathbf{R}} \frac{e^{ik_\omega |\mathbf{r}-\mathbf{R}|}}{|\mathbf{r}-\mathbf{R}|}, \tag{A 1}$$

where $\mathbf{R}$ ($\mathbf{r}$) denote the displacement vectors of real lattice sites(field point), $k_\omega = \omega/c$, and $\mathbf{k}_B$ is the Bloch wave vector.

The lattice sum is related to the limit of above series by

$$S(k_\omega, \mathbf{k}_B) \equiv \sum_{\mathbf{R}}{}' e^{i\mathbf{k}_B \cdot \mathbf{R}} \frac{e^{ik_\omega R}}{R} = \lim_{r \to 0} Q(\mathbf{r}, k_\omega, \mathbf{k}_B) - \lim_{z \to 0_+} \frac{e^{ik_\omega z}}{z}, \tag{A 2}$$

where $\sum'$ means a summation excluding the origin, and the $\lim_{z \to 0_+} \frac{e^{ik_\omega z}}{z}$ is the self-radiated field of the reference particle at the origin.

When the summation over real lattice is to be done on 2D array, we know from Poisson's summation formula[30] that

$$Q(\mathbf{r}, k_\omega, \mathbf{k}_B) = \sum_{\mathbf{G}} \frac{1}{\Omega} F(\mathbf{k}_B + \mathbf{G}) e^{i(\mathbf{k}_B + \mathbf{G}) \cdot \boldsymbol{\rho}}, \tag{A 3}$$

where the $\Omega = |\mathbf{a} \times \mathbf{b}|$ is the area of the unit cell of real 2D lattice, $\mathbf{G}$ are the displacement vectors of 2D reciprocal lattice sites, and $F(\mathbf{k})$ is the Fourier transform of function $e^{ik_\omega r}/r$,

$$F(\mathbf{k}) = \int d\boldsymbol{\rho} \, e^{-i\mathbf{k} \cdot \mathbf{r}} \frac{e^{ik_\omega r}}{r} = \frac{2\pi i e^{ik_z z}}{k_z}, \tag{A 4}$$

$$k_z = \sqrt{k_\omega^2 - k^2}. \tag{A 5}$$

Here $\mathbf{r} = \boldsymbol{\rho} + z\hat{z}$ are 3D vectors, while $\boldsymbol{\rho}$, $\mathbf{R}$, $\mathbf{G}$ and $\mathbf{k}_B$ are 2D vectors.

Putting Eq. (A4), (A3) into (A2), we have

$$S(k_\omega, \mathbf{k}_B) = \lim_{z \to 0_+} \left( \frac{2\pi i}{\Omega} \sum_{\mathbf{G}} \frac{e^{ik_z(\mathbf{G})z}}{k_z(\mathbf{G})} - \frac{e^{ik_\omega z}}{z} \right), \tag{A 6}$$

$$k_z(\mathbf{G}) = \sqrt{k_\omega^2 - |\mathbf{k}_B + \mathbf{G}|^2}. \tag{A 7}$$

Given the case $\mathbf{k}_B = 0$, the real part of $S$ is found to have finite limit when $k_\omega$ and $z$ are tending to zero together. That limit can be defined as a new constant

$$D = \operatorname{Re}[S(0_+, 0)] = \lim_{k_\omega \to 0_+} \operatorname{Re}\left( \sum_{\mathbf{R}}{}' \frac{e^{ik_\omega R}}{R} \right) = \lim_{z \to 0_+} \left( \frac{2\pi}{\Omega} \sum_{\mathbf{G}}{}' \frac{e^{-zG}}{G} - \frac{1}{z} \right). \tag{A 8}$$

Adding to and subtracting from the right part of (A6) the value $D$ and taking (A8) into account, we have

$$S(k_\omega, \mathbf{k_B}) = D + \frac{2\pi i}{\Omega} \frac{1}{\sqrt{k_\omega^2 - k_B^2}} + \lim_{z \to 0_+} \left( \frac{1}{z} - \frac{e^{ik_\omega z}}{z} \right)$$

$$+ \frac{2\pi}{\Omega} \lim_{z \to 0_+} \left[ \sum_{\mathbf{G}}{}' \frac{e^{-z\sqrt{|\mathbf{k_B} + \mathbf{G}|^2 - k_\omega^2}}}{\sqrt{|\mathbf{k_B} + \mathbf{G}|^2 - k_\omega^2}} - \frac{e^{-zG}}{G} \right] \quad \text{(A 9)}$$

$$= D + \frac{2\pi i}{\Omega} \frac{1}{\sqrt{k_\omega^2 - k_B^2}} - ik_\omega + \frac{2\pi}{\Omega} \sum_{\mathbf{G}}{}' \left( \frac{1}{\sqrt{|\mathbf{k_B} + \mathbf{G}|^2 - k_\omega^2}} - \frac{1}{G} \right).$$

In Fig. 8, we compare the Simovski's series with a brute-force real-space sum. While the original sum does not seem to converge, the Simovski's series quickly converges in a small circle with radius of a few hundred reciprocal-lattice constant. The convergence of the original sum becomes better in $k_B > \omega/c$ region than in $k_B < \omega/c$, because two neighboring dipoles are approximately anti-phase when $k_B > \omega/c$ such that every two dipoles together form a quadruple of which the lattice sum has good convergence.

**Appendix B.**

Let

$$\sum_{\mathbf{R}\neq 0}\left(\frac{1}{R^3}-\frac{ik_\omega}{R^2}\right)e^{i(k_\omega R+\mathbf{k_B}\cdot\mathbf{R})}=\frac{1}{\Omega}\int_{R_{min}}^{\infty}\left(\frac{1}{r^2}-\frac{ik_\omega}{r}\right)e^{ik_\omega r}dr\int_0^{2\pi}e^{ik_B r\cos\theta}d\theta$$

$$=\frac{2\pi}{\Omega}\int_{R_{min}}^{\infty}\left(\frac{1}{r^2}-\frac{ik_\omega}{r}\right)e^{ik_\omega r}J_0(k_B r)dr=\frac{2\pi}{\Omega}I_1$$

(B 1)

(Here we also assume that the lattice sum is exactly equivalent to the integral from $R_{min}$ to infinity, while $R_{min}$ is unknown initially)

where

$$I_1=\int_{R_{min}}^{\infty}\left(\frac{1}{r^2}-\frac{ik_\omega}{r}\right)e^{ik_\omega r}J_0(k_B r)dr=-\int_{R_{min}}^{\infty}J_0(k_B r)d\frac{e^{ik_\omega r}}{r}$$

$$=J_0(k_B r)\frac{e^{ik_\omega r}}{r}\bigg|_\infty^{R_{min}}+\int_{R_{min}}^{\infty}\frac{e^{ik_\omega r}}{r}dJ_0(k_B r)$$

$$=\frac{J_0(k_B R_{min})}{R_{min}}e^{ik_\omega R_{min}}-k_B^2\int_{R_{min}}^{\infty}\frac{J_1(k_B r)}{k_B r}e^{ik_\omega r}dr,$$

(B 2)

$$\int_{R_{min}}^{\infty}\frac{J_1(k_B r)}{k_B r}e^{ik_\omega r}dr=\frac{i}{k_\omega+\sqrt{k_\omega^2-k_B^2}}-\int_0^{R_{min}}\frac{J_1(k_B r)}{k_B r}e^{ik_\omega r}dr,$$

(B 3)

and

$$\int_0^{R_{min}}\frac{J_1(k_B r)}{k_B r}e^{ik_\omega r}dr\approx\frac{i}{2k_\omega}\left(1-e^{ik_\omega R_{min}}\right).$$

(B 4)

Finally we have

$$\therefore I_1=\frac{J_0(k_B R_{min})}{R_{min}}e^{ik_\omega R_{min}}-k_B^2\left(\frac{i}{k_\omega+\sqrt{k_\omega^2-k_B^2}}-\int_0^{R_{min}}\frac{J_1(k_B r)}{k_B r}e^{ik_\omega r}dr\right)$$

$$=\frac{J_0(k_B R_{min})}{R_{min}}e^{ik_\omega R_{min}}-k_B^2\left[\frac{i}{k_\omega+\sqrt{k_\omega^2-k_B^2}}-\frac{i}{2k_\omega}\left(1-e^{ik_\omega R_{min}}\right)\right].$$

(B 5)

The parameter $R_{min}$ can be determined in the electrostatic limit, by making $k_B=0$ and $k_\omega\to 0$. Then we have

$$\sum_{\mathbf{R}\neq 0}\frac{1}{R^3}=\frac{2\pi}{\Omega}\lim_{k_\omega\to 0}I_1(k_\omega,0)=\frac{2\pi}{\Omega}\frac{1}{R_{min}}.$$

(B 6)

For square lattice, the numerical result is $R_{min}=d/1.438$.

Similarly, let

$$\sum_{\mathbf{R}\neq 0}\left(\frac{k_\omega^2}{R}\right)e^{i(k_\omega R+\mathbf{k_B}\cdot\mathbf{R})}=\frac{1}{\Omega}\int_{R_{\min}}^{\infty}k_\omega^2 e^{ik_\omega r}dr\int_0^{2\pi}e^{ik_B r\cos\theta}d\theta$$

$$=\frac{2\pi}{\Omega}\int_{R_{\min}}^{\infty}k_\omega^2 e^{ik_\omega r}J_0(k_B r)dr=\frac{2\pi}{\Omega}I_2,\qquad\text{(B 7)}$$

where

$$I_2=\int_{R_{\min}}^{\infty}k_\omega^2 e^{ik_\omega r}J_0(k_B r)dr=-ik_\omega\int_{R_{\min}}^{\infty}J_0(k_B r)d\left(e^{ik_\omega r}\right)$$

$$=-ik_\omega J_0(k_B r)e^{ik_\omega r}\Big|_{R_{\min}}^{\infty}+ik_\omega\int_{R_{\min}}^{\infty}e^{ik_\omega r}d\left[J_0(k_B r)\right]\qquad\text{(B 8)}$$

$$=ik_\omega\left[J_0(k_B R_{\min})e^{ik_\omega R_{\min}}-k_B\int_{R_{\min}}^{\infty}J_1(k_B r)e^{ik_\omega r}dr\right],$$

$$\int_{R_{\min}}^{\infty}J_1(k_B r)e^{ik_\omega r}dr=\frac{-k_B}{\sqrt{k_\omega^2-k_B^2}\left(k_\omega+\sqrt{k_\omega^2-k_B^2}\right)}-\int_0^{R_{\min}}J_1(k_B r)e^{ik_\omega r}dr,\qquad\text{(B 9)}$$

and

$$\int_0^{R_{\min}}J_1(k_B r)e^{ik_\omega r}dr\approx\frac{k_B}{2k_\omega^2}\left[\left(1-ik_\omega R_{\min}\right)e^{ik_\omega R_{\min}}-1\right].\qquad\text{(B 10)}$$

In (B4) and (B10) the integrands are approximated as $\frac{J_1(k_B r)}{k_B r}\approx\frac{1}{2}$ and $J_1(k_B r)\approx\frac{k_B r}{2}$, the same as those in Ref. 19.

Therefore

$$\beta_T=\sum_{\mathbf{R}\neq 0}\left(-\frac{1}{R^3}+\frac{ik_\omega}{R^2}+\frac{k_\omega^2}{R}\right)e^{i(k_\omega R+\mathbf{k_B}\cdot\mathbf{R})}=\frac{2\pi}{\Omega}(I_2-I_1).\qquad\text{(B 11)}$$

By approximating the lattice sum for dense dipole array as integrals, we have

$$\beta_{L,I}=\frac{1}{\Omega}\int_{R_{\min}}^{\infty}\int_0^{2\pi}e^{ik_B r\cos\theta}e^{ik_\omega r}\left[(3\cos^2\theta-1)\left(\frac{1}{r^2}-\frac{ik_\omega}{r}\right)+k_\omega^2\sin^2\theta\right]drd\theta,\qquad\text{(B 12)}$$

$$\beta_{T,I}=\frac{1}{\Omega}\int_{R_{\min}}^{\infty}\int_0^{2\pi}e^{ik_B r\cos\theta}e^{ik_\omega r}\left[(3\sin^2\theta-1)\left(\frac{1}{r^2}-\frac{ik_\omega}{r}\right)+k_\omega^2\cos^2\theta\right]drd\theta.\qquad\text{(B 13)}$$

From Ref. 19, we know that $\beta_{L,I}=2\pi(I_1-I_3)/\Omega$ and $\beta_{T,I}=2\pi(I_2+I_3)/\Omega$, where

$$I_3=\left(\frac{1}{R_{\min}}-ik_\omega\right)\frac{J_1(k_B R_{\min})}{k_B R_{\min}}e^{ik_\omega R_{\min}}.\qquad\text{(B 14)}$$

Then for $k_B < k_\omega$, we have

$$\mathrm{Re}[I_1] = \frac{J_0(k_B R_{\min})}{R_{\min}} \cos(k_\omega R_{\min}) + \frac{k_B^2}{2k_\omega} \mathrm{Re}\left[i\left(1 - e^{ik_\omega R_{\min}}\right)\right]$$
$$= \frac{J_0(k_B R_{\min})}{R_{\min}} \cos(k_\omega R_{\min}) + \frac{k_B^2}{2k_\omega} \sin(k_\omega R_{\min}),$$

(B 15)

$$\mathrm{Re}[I_2] = -k_\omega \mathrm{Im}\left[J_0(k_B R_{\min}) e^{ik_\omega R_{\min}} - k_B \int_{R_{\min}}^{\infty} J_1(k_B r) e^{ik_\omega r} dr\right]$$
$$= -k_\omega \left\{J_0(k_B R_{\min}) \sin(k_\omega R_{\min}) - k_B \mathrm{Im}\left[\frac{-k_B}{\sqrt{k_\omega^2 - k_B^2}\left(k_\omega + \sqrt{k_\omega^2 - k_B^2}\right)} - \int_0^{R_{\min}} J_1(k_B r) e^{ik_\omega r} dr\right]\right\}$$
$$= -k_\omega \left\{J_0(k_B R_{\min}) \sin(k_\omega R_{\min}) + \frac{k_B^2}{2k_\omega^2} \mathrm{Im}\left[(1 - ik_\omega R_{\min}) e^{ik_\omega R_{\min}} - 1\right]\right\}$$
$$= -k_\omega \left\{J_0(k_B R_{\min}) \sin(k_\omega R_{\min}) + \frac{k_B^2}{2k_\omega^2}\left[\sin(k_\omega R_{\min}) - k_\omega R_{\min} \cos(k_\omega R_{\min})\right]\right\},$$

(B 16)

$$\mathrm{Re}[I_3] = \frac{J_1(k_B R_{\min})}{k_B R_{\min}} \mathrm{Re}\left[\left(\frac{1}{R_{\min}} - ik_\omega\right) e^{ik_\omega R_{\min}}\right]$$
$$= \frac{J_1(k_B R_{\min})}{k_B R_{\min}} \left[\frac{\cos(k_\omega R_{\min}) + k_\omega R_{\min} \sin(k_\omega R_{\min})}{R_{\min}}\right].$$

(B 17)

Noting that the Bessel functions $J_0$ and $J_1$ in the integrals can be expanded in polynomials for small arguments as $J_0(x) \sim 1 - x^2/4 + O(x^4)$ and $J_1(x)/x \sim 1/2 - x^2/16 + O(x^4)$, and we have $k_B R_{\min} \ll 1$ inside the light-cone, the form of $\mathrm{Re}[\beta_\sigma]$ near resonant frequencies $\omega_\sigma$ can be approximated in

$$\mathrm{Re}[\beta_T] = \frac{2\pi}{\Omega}\left\{-\frac{k_B^2}{2k_\omega}\left[2\sin(k_\omega R_{\min}) - k_\omega R_{\min} \cos(k_\omega R_{\min})\right] - \frac{J_0(k_B R_{\min})}{R_{\min}}\left[\cos(k_\omega R_{\min}) + k_\omega R_{\min} \sin(k_\omega R_{\min})\right]\right\}$$
$$\approx \frac{2\pi}{\Omega}\left\{-\frac{k_B^2}{2k_\omega}\left[2\sin(k_\omega R_{\min}) - k_\omega R_{\min} \cos(k_\omega R_{\min})\right] - \frac{4 - (k_B R_{\min})^2}{4R_{\min}}\left[\cos(k_\omega R_{\min}) + k_\omega R_{\min} \sin(k_\omega R_{\min})\right]\right\}$$
$$\approx \frac{2\pi}{\Omega}\left[-\frac{\cos(k_\omega R_{\min}) + k_\omega R_{\min} \sin(k_\omega R_{\min})}{R_{\min}} + k_B^2 R_{\min}^2 \frac{3\cos(k_\omega R_{\min}) - 3\sin(k_\omega R_{\min})}{4R_{\min}}\right]$$

(B 18)

$$\text{Re}[\beta_{TI}] = \frac{2\pi}{\Omega} \left\{ \frac{J_1(k_B R_{\min})}{k_B R_{\min}} \left[ \frac{\cos(k_\omega R_{\min}) + k_\omega R_{\min} \sin(k_\omega R_{\min})}{R_{\min}} \right] \right.$$

$$\left. - J_0(k_B R_{\min}) k_\omega \sin(k_\omega R_{\min}) - \frac{k_B^2}{2k_\omega} \left[ \sin(k_\omega R_{\min}) - k_\omega R_{\min} \cos(k_\omega R_{\min}) \right] \right\}$$

$$\approx \frac{2\pi}{\Omega} \left\{ \left( \frac{1}{2} - \frac{k_B^2 R_{\min}^2}{16} \right) \left[ \frac{\cos(k_\omega R_{\min}) + k_\omega R_{\min} \sin(k_\omega R_{\min})}{R_{\min}} \right] \right.$$

$$\left. - \left( 1 - \frac{k_B^2 R_{\min}^2}{4} \right) \frac{k_\omega R_{\min} \sin(k_\omega R_{\min})}{R_{\min}} - \frac{k_B^2 R_{\min}^2}{2k_\omega R_{\min}} \left[ \frac{\sin(k_\omega R_{\min}) - k_\omega R_{\min} \cos(k_\omega R_{\min})}{R_{\min}} \right] \right\}$$

$$\approx \frac{2\pi}{\Omega} \left[ \frac{\cos(k_\omega R_{\min}) - k_\omega R_{\min} \sin(k_\omega R_{\min})}{2R_{\min}} + k_B^2 R_{\min}^2 \frac{7\cos(k_\omega R_{\min}) - 5\sin(k_\omega R_{\min})}{16 R_{\min}} \right]$$

(B 19)

$$\text{Re}[\beta_{LI}] = \frac{2\pi}{\Omega} \left\{ \frac{J_0(k_B R_{\min})}{R_{\min}} \cos(k_\omega R_{\min}) + \frac{k_B^2}{2k_\omega} \sin(k_\omega R_{\min}) \right.$$

$$\left. - \frac{J_1(k_B R_{\min})}{k_B R_{\min}} \left[ \frac{\cos(k_\omega R_{\min}) + k_\omega R_{\min} \sin(k_\omega R_{\min})}{R_{\min}} \right] \right\}$$

$$\approx \frac{2\pi}{\Omega} \left\{ \frac{4 - (k_B R_{\min})^2}{4 R_{\min}} \cos(k_\omega R_{\min}) + \frac{k_B^2 R_{\min}^2}{2k_\omega R_{\min}} \frac{\sin(k_\omega R_{\min})}{R_{\min}} \right.$$

$$\left. - \left( \frac{1}{2} - \frac{k_B^2 R_{\min}^2}{16} \right) \left[ \frac{\cos(k_\omega R_{\min}) + k_\omega R_{\min} \sin(k_\omega R_{\min})}{R_{\min}} \right] \right\}$$

$$\approx \frac{2\pi}{\Omega} \left[ \frac{\cos(k_\omega R_{\min}) - k_\omega R_{\min} \sin(k_\omega R_{\min})}{2R_{\min}} + k_B^2 R_{\min}^2 \frac{-3\cos(k_\omega R_{\min}) + 9\sin(k_\omega R_{\min})}{16 R_{\min}} \right]$$

(B 20)

where we have taken the approximation $k_\omega R_{\min} = \frac{\omega R_{\min}}{c} \approx \frac{\omega_0 R_{\min}}{c} \approx 1$ near resonance, in order to simplify the coefficient associated with the quadratic term $k_B^2 R_{\min}^2$, and such simplified coefficient can make it easier to compare between different modes.